\documentstyle[12pt]{article}

\newcommand{\be}{\begin{equation}}
\newcommand{\ee}{\end{equation}}

\newcommand{\bea}{\begin{eqnarray}}
\newcommand{\eea}{\end{eqnarray}}

\begin{document}
\title{Classical Kinetic Theory of Landau Damping\\ 
for Self-interacting Scalar Fields\\
 in the Broken Phase} 

\author{{A. Patk\'os$^{1}$ and Zs. Sz\'ep$^{2}$}\\
{Department of Atomic Physics}\\
{E\"otv\"os University, Budapest, Hungary}\\
}
\vfill

\footnotetext[1]{patkos@ludens.elte.hu}
\footnotetext[2]{szepzs@hercules.elte.hu}
\maketitle
\begin{abstract}
The classical kinetic theory of one-component self-interacting sca\-lar fields 
is formulated in the broken symmetry phase and applied to the phenomenon 
of Landau damping. The domain
of validity of the classical approach is found by comparing with the result
of a 1-loop quantum calculation.
\end{abstract}
\section{Introduction}
For high enough temperature leading non-equlibrium transport effects
of large wave number $|{\bf k}|>>T$ fluctuations
can be reproduced by a kinetic theory of the corresponding response 
functions. This approach has been applied to the features of 
the QCD plasma \cite{heinz86} and
proved successfull in reproducing the contribution of hard loops to the
Green functions of low $|{\bf k}|<<T$ modes in Abelian and non-Abelian
gauge theories \cite{liu94,iancu94}.

These investigations have assumed tacitly that no
background fields are present, all symmetries are restored. This assumption
is certainly not valid for all field theoretical models, since in some scalar
models with specific internal symmetries one finds arguments in favor of 
non-restoration of the symmetry at arbitrary high temperature \cite{weinb74,
skage85}. Also certain approaches to non-Abelian
gauge+Higgs systems provide evidence for non-zero scalar expectation
values even in the high-temperature phase \cite{buchm95}. Therefore, there
exist intuitive hints for possible relevance of the classical kinetic
considerations in a non-zero scalar background even at high temperature.

The classical kinetic theory for self-interacting scalar fields has been
derived first by Danielewicz and Mr\'owczynski \cite{mrow90} and its features
are still being discussed  \cite{brandt98}. These papers deal with scalar 
theories of non-negative squared mass parameter. Therefore their
results are useful for the calculation of transport characteristics
(plasmon frequency, damping rates) in the restored symmetry phase. 
In Ref. \cite{brandt98} an effective action is derived, which accounts for 
the loop contribution of high-$k$ fluctuations to the Green functions of the 
low-$k$ modes. As a parallel evolution one may note that a classical cut-off 
field theoretic approach to time dependent 
phenomena in the symmetric phase has been developed in Refs. \cite{aarts97,  
aarts98,buchm97,buchm98}. Quantitative results relevant for quantum 
systems were obtained by matching the classical theory to the parameters of 
the quantum theory. This 
effective theory reproduces, for instance, correctly the on-shell damping 
rate in the high temperature phase.

The present note rederives in a fully relativistic Lagrangian 
formalism the kinetic theory of scalar fields for the case of non-zero 
background field (generated by a negative 
mass squared parameter). It starts by proposing a Lagrangian
for the effective particles, which accounts for the effect of the high-k 
$\varphi$ modes on the low-k fluctuations. A detailed  motivation for 
this Lagrangian is provided by recalling results of earlier investigations.
An advantage of this proposition is that it leads to the
induced source density of the low-k fluctuations directly, without any 
reference to the quantum theory.
Evidence for the correctness of the effective lagrangian can be presented
by comparing its consequences with the results of the corresponding 
quantum calculations. As a first test we compute a physically meaningful 
quantity, the Landau damping coefficient for the off-shell scalar 
fluctuations. This effect is present only in the broken symmetry phase
of the scalar theory. Its classical value is compared with the
result of a 1-loop quantum calculation \cite{boya96}, establishing in this way
the domain of validity of the proposed classical treatment.

\section{Kinetic theory of the high-k modes in a non-zero background}

The effective gas of high-frequency fluctuations is out of thermal
equilibrium if an inhomogenous low frequency background fluctuation is present.
This state of the gas induces a source term into the wave equation of the 
low-k modes. A unified description of the two effects can be attempted if 
in addition to the action $S_{cl}$ describing the low-frquency dynamics
an appropriate Lagrangian can be introduced for a gas particle coupled to the
low-frequency field along its trajectory $\xi_\mu (\tau )$:
\bea
&
S_{eff}=S_{cl}[\varphi ]+\Delta S,\nonumber\\
&
\Delta S=\int d\tau L_{particle} (\varphi (\xi (\tau ))).
\label{action0}
\eea
Variation of this additional piece of action with respect to the 
field variable $\varphi(x)$ should yield the induced source term to the wave
equation of the low  frequency modes when averaged over the statistical 
distribution of the gas particles. The distribution can be derived from
the solution of a Boltzmann equation describing the gas in the background
field $\varphi$. Variation of (\ref{action0}) with respect to the particle
trajectory provides the expression of the force exerted on the particle by the
external field $\varphi$. This information is used in the collisionless
Boltzmann-equation.

The effective particle Lagrangian for the one-component selfinteracting scalar 
fields described by the theory:
\be
L[\varphi ]={1\over 2}(\partial_\mu\varphi )^2-{1\over 2}m^2\varphi^2-
{\lambda\over 24}\varphi^4
\ee
can be guessed intuitively by making use of the equation of motion for the 
real time Green function: $\Delta^{>}(x,y)=\langle\varphi (x)\varphi (y)
\rangle$. Its Wigner-transform in the collisionless case
fulfills \cite{mrow90}
\be
\bigl(p_{\mu}{\partial\over\partial X_\mu}+{1\over 2}{\partial m^2(X)\over
\partial X_\mu}{\partial\over\partial p_\mu}\bigr)\Delta^{>}(X,p)=0,\ee
where $m^2(X)=m^2+(\lambda /2)\varphi^2(X)$, with $\varphi$ representing 
the background at $X=(x+y)/2$. This equation suggests the relation
\be 
mF_\mu ={1\over 2}\partial_\mu m^2(X),
\label{force0}
\ee
which is equivalent to 
\be
L_{particle}=-m[\varphi (x)].
\label{particle}
\ee

On the basis of this argument we propose
for the classical effective action of an effective particle representation 
of the high-$k$ 
modes of the one-component, self-interacting scalar field
\bea
&
\Delta S=-\int M_{loc}[\bar\varphi ,\varphi (\xi (\tau))]d\tau,\nonumber\\
&
M_{loc}^2[\bar\varphi ,\varphi (x)]=m^2+{\lambda\over 2}
(\bar\varphi +\varphi (x))^2,
\label{action}
\eea
where $m^2$ is of negative sign, $\bar\varphi$ denotes the spontanously
generated non-zero average
background field, and $\varphi (x)$ is the amplitude of the low-$k$ 
fluctuation field at $x$ in the broken phase. 

We propose to write the Boltzmann-equation for the effective particle with
an $x$-independent mass $m_{eff}$ (see below), which leads to a slight 
deviation from the kinetic 
equation derived by \cite{mrow90}. Its advantage is that the momentum variable
of the one-particle distribution is defined through an $x$-independent 
relation.  This definition gives slightly different ${\cal O}(\lambda^2)$ 
thermal
mass to the low-frequency waves, but the damping coefficient remains 
unchanged. The necessary input into the derivation of the Boltzmann-equation 
is the equation for the kinetic momentum:
\be
p_\mu =m_{eff}{d\xi\over d\tau},\quad m^2_{eff}(\bar\varphi )=m^2
+{\lambda\over 2}\bar\varphi^2.
\ee
The relevant equation can be found from the canonical (Euler-Lagrange) 
equation:
\bea
&
{d\over d\tau}(M_{loc}[\bar\varphi  ,\varphi (x)]\dot\xi_\mu )=
{dM_{loc}[\bar\varphi ,\varphi (x)]\over dV[\bar\varphi ,\varphi (x)]}
\partial_\mu V[\bar\varphi ,\varphi (x)],\nonumber\\
&
V[\bar\varphi ,\varphi (x)]=\lambda(\bar\varphi\varphi+{1\over 2}
\varphi^2), \quad x=\xi (\tau ).
\eea
From this equation, using explicitly the definition of $M_{loc}$, one can 
express the proper-time derivative of the kinetic
momentum.
\be
m_{eff}{dp_\mu\over d\tau}={m_{eff}^2\over 2M_{loc}^2}
\biggl[\partial_\mu V[\bar\varphi ,\varphi (x)]-
{p_\mu\over m^2_{eff}}(p\cdot\partial )V[\bar\varphi ,\varphi ]\biggr].
\label{force}
\ee
The second term on the right hand side is missing from (\ref{force0}).

With help of (\ref{force}) one writes the collisionless
Boltzmann-equation for the gas of these particles:
\be
(p\cdot\partial )f(x,p)+{m_{eff}^2(\bar\varphi )\over
2M_{loc}^2(\varphi ,\bar\varphi )}
\biggl[\partial_\mu V[\bar\varphi ,\varphi (x)]-
{p_\mu\over m^2_{eff}}(p\cdot\partial )V[\bar\varphi ,\varphi ]\biggr]
{\partial f(x,p)\over\partial p_\mu}=0.
\label{boltz}
\ee
Its perturbative solution in the weak coupling limit 
$\lambda <<1$ is searched for in the form:
\be
f(x,p)=f_0(p)+\lambda f_1(\varphi (x), p), \quad f_0(p)={1\over
e^{\beta p_0}-1}.
\ee
(One may emphasize that the small parameter allowing iterative 
solution of Eq.(\ref{boltz}) is hidden in $V$).
The formal solution is easily found in full structural agreement with the 
result of \cite{brandt98}:
\be
f_1(\varphi (x),p)=-{1\over 2}\biggl[{1\over (p\cdot\partial )}(\bar\varphi+
\varphi (x))\partial_\mu \varphi (x)-{p_\mu\over m^2_{eff}}
(\bar\varphi\varphi (x) +{1\over 2}\varphi^2(x))\biggr]{df_0\over dp_\mu}.
\ee

\section{Source density induced by field fluctuations}

The variation of (\ref{action}) with respect to $\varphi$ provides
the induced source term for the low-$k$ fluctuations.
For a number of particles moving on definite trajectories one has
\be
j(x)={1\over 2}\sum_i\int d\tau\biggl({1\over M_{loc}[\bar\varphi ,\varphi ]}
{dV[\bar\varphi ,\varphi (x)]\over d\varphi}\biggr)
\delta^{(4)}(x-\xi_i(\tau )).
\ee
Here the index $i$ refers to the trajectory of the $i$-th particle.
Statistical average over the full momentum space and in a small volume
around the point $x$ introduces the one-particle distribution
\be
m_{eff}\int {d^3p\over (2\pi )^3p_0}f(x,p)=
\langle\int d\tau\sum_i\delta^{(4)}({\bf x}-{\bf\xi}_i(\tau ))\rangle
\ee
and  produces the following "macroscopic" source density:
\be
j_{av}(x)={1\over 2}\int {d^3p\over (2\pi )^3p_0}{m_{eff}(\bar\varphi )\over
M_{loc}[\bar\varphi ,\varphi ]}{dV[\bar\varphi ,\varphi ]\over d\varphi }
f(x,p), \quad p_0^2=m_{eff}^2+{\bf p}^2.
\label{source0}
\ee

The induced linear response to $\varphi (x)$ is found by retaining terms with
linear functional dependence of $j_{av}$ on $\varphi$. The contribution 
of the equilibrium distribution $f_0$ is easily found:
\be
j_{av}^{(0)}(x)={\lambda\over 2}\varphi (x)\int{d^3p\over (2\pi )^3p_0}
f_0(p_0)\biggl(1-{\lambda\bar\varphi^2\over 2m_{eff}^2}\biggr).
\label{source}
\ee
The ${\cal O}(\lambda^2)$ contribution is the result of the expansion of
$M_{loc}$ in the denominator of (\ref{source0}). This contribution to the 
thermal mass in the high-$T$ limit gives the correct limiting  
value for the one-component scalar theory.

The terms induced by $f_1$ give rise to the following expression linear in
$\varphi$:
\bea
&
j_{av}^{(1)}(x)={\lambda^2\bar\varphi\over 2}\int {d^3p\over (2\pi )^3p_0}
f_1(\varphi (x), p)\nonumber\\
&
\sim -{\lambda^2\bar\varphi^2\over 4}\int {d^3p\over (2\pi )^3p_0}{1\over 
(p\cdot\partial )} \partial_0\varphi (x){df_0\over dp_0}+{\lambda^2\bar
\varphi^2\varphi (x)\over 8\pi^2}\int_{m_{eff}}^\infty dp_0f_0(p_0)
{1\over\sqrt{p_0^2-m_{eff}^2}}.
\label{resp}
\eea
The damping effect arises from the first term, while from the second another
${\cal O}(\lambda^2)$ contribution is obtained to the thermal mass.

The imaginary part of the linear response can be evaluated by the use of the
principal value theorem, when the $\epsilon$-prescription of Landau is 
applied to the $(p\cdot\partial )^{-1}$ operator in the first term on
the right hand side of (\ref{resp}). 
For an explicit expression one may assume that
$\varphi$ represents an off-mass-shell fluctuation characterised by the
4-vector: $(\omega ,{\bf k})$, that is $\varphi (x)=\varphi (k\cdot x)$. 
The evaluation of the integral is a not too difficult excercise. 
One is led to the following expression for the imaginary part of the
linear response function:
\be
{\rm Im}\Sigma =
{\lambda^2\bar\varphi^2\over 16\pi}
{1\over e^{\beta m_eff/\sqrt{1-\omega^2/|{\bf k}|^2}}-1}\cdot
{\omega\over |{\bf k}|}\Theta (|{\bf k}|^2-\omega^2).
\label{damp}
\ee

\section{Discussion}

In this note we have shown that in the broken phase of scalar theories 
Landau-type damping phenomenon occurs according to an effective 
classical kinetic theory. The clue to its existence is the presence of the 
$\sim\bar\varphi\varphi$ term in the fluctuating part of the local mass, which
is responsible for the emergence of a linear source-amplitude relation with
non-zero imaginary part. 
On the other hand the imaginary part of the self-energy has been 
computed earlier in the $\varphi^4$-theory using 
its real time quantum formulation \cite{boya96}, where 
``a term reminiscent of Landau damping'' was found:
\bea
&
{\rm Im}\Sigma_{Landau}={\lambda^2\bar\varphi^2T\over 16\pi |{\bf k}|}
\ln{1-e^{-\beta\omega_k^+}\over 1-e^{-\beta\omega_k^-}}
\Theta (|{\bf k}|^2-\omega^2),\nonumber\\
&
\omega_{\bf k}^{\pm}=\left|{\omega\over 2}\pm {|{\bf k}|\over 2}
\sqrt{1-{4M^2\over \omega^2-|{\bf k}|^2}}\right|.
\label{dan}
\eea
Here $M$ is the temperature dependent mass parameter in the finite 
temperature quantum theory. With explicit calculation one may check that
in the limit $\beta\omega ,\beta |{\bf k}|<<1$ the expression of the
damping rate calculated from (\ref{dan}) goes over into (\ref{damp}), under 
the assumption that $m_{eff}$ of the classical theory is identified with $M$. 
In this way the classical theory provides a convincing argument 
clearly demonstrating that Landau damping is the correct interpretation of 
the result derived from quantum theory. 

The extension of our discussion to the $n$-component scalar fields 
in the broken phase forces us to accept 
the choice (\ref{force0}), since for the Goldstone fields $m_{eff}=0$.
Then one can see by analysing the equations of motion of the
Green-functions composed from the different field components, 
that with accuracy 
${\cal O}(\lambda^2)$ one-particle distributions for each type of particles 
have their separate kinetic equation, decoupled from each other. 
Landau damping of the heavy (``Higgs'') mode
can be described in full analogy with the treatment of the present paper,
writing for each particle a mechanical action of the form (\ref{action}) with 
appropriate local mass expressions. Damping of the Goldstone modes is the 
result of the non-zero correlation between these and the ``Higgs'' modes.
One can construct an additional non-local, velocity dependent mechanical 
action for the heavy particles describing this effect \cite{jako98}.

If one completes the kinetic theory of the pure non-Abelian gauge fields 
by appropriate scalar fields, an important generalisation of the
present discussion can be made to the 
damping rates of gauge+Higgs systems, where also the interpretation 
of lattice simulations \cite{tang98} represents a non-trivial challenge 
\cite{bode98}. Some results concerning the hard thermal loop effects on 
thermal masses of the gauge bosons
have been obtained recently in \cite{man98}. An approach based on  the 
coupled kinetic system is presently under study \cite{jako98}.

The authors acknowledge informative discussions on the subject of the paper
with A. Jakov\'ac and P. Petreczky. This work has been supported by the 
research contract OTKA-T22929.

\end{document}